\documentclass[12pt,preprint]{aastex}
\usepackage{apjfonts}
\usepackage{graphicx}
\usepackage{lineno}
\usepackage{ulem}

\begin{document}

\title{KIC 8840638: A new eclipsing binary consisting of $\delta$ Scuti-type oscillations with an extremely cold companion star}
\shorttitle{KIC 8840638: A Eclipsing Binary Consisting of a $\delta$ Scuti star}

\author{
Tao-Zhi Yang\altaffilmark{1},
Zhao-Yu Zuo\altaffilmark{1},
Jun-Hui Liu\altaffilmark{2},
Xiao-Ya Sun\altaffilmark{1}
}

\altaffiltext{1}{
School of Physics, Xi'an Jiaotong University, 710049 Xi'an, People's Republic of China; e-mail:zuozyu@xjtu.edu.cn}

\altaffiltext{2}{
Department of Astronomy, Xiamen University, 361005 Xiamen, Fujian, People's Republic of China.}


\begin{abstract}

In this paper, we analyze the light variation of KIC 8840638 using high-precision time-series data from $Kepler$ mission. The analysis reveals this target is a new Algol-type eclipsing binary system with a $\delta$ Scuti component, not a pure single $\delta$ Scuti star previously known. The frequency analysis of the short-cadence light curve reveals 95 significant frequencies, most of which lie in a frequency range of 23$-$32 d$^{-1}$. Among them, seven independent frequencies are detected in the typical frequency range of $\delta$ Scuti stars and they are identified as pressure modes. Besides, the orbital frequency $f_{orb}$ (=0.320008 d$^{-1}$) and its harmonics are also detected directly in the frequency spectrum. The binary modellings derived from the Wilson-Devinney code indicate the binary parameters are not affected by the pulsations, and the binary system is in semi-detached configure with a mass ratio of $q$ = 0.454, an inclination angle of 58 degrees, and a temperature difference of larger than 4000 K between the two components. The derived parameters suggest the primary of this system is a main-sequence star with spectral type about A7V, while the secondary seems to be the coolest companion star among the known semi-detached Algol systems, implying it may have stepped into a highly evolved stage.

\end{abstract}


\section{Introduction}

Eclipsing binaries (EBs) are primary targets for us to obtain the precise fundamental stellar parameters \citep{1980ARA&A..18..115P,2010A&ARv..18...67T}. With high-precision time-series photometric and spectroscopic observations, the mass and radius of each component in EBs can be determined precisely within 1 percent \citep{2005MNRAS.363..529S,2008A&A...487.1095C}. Additionally, with high-precise and long-term photometric time series, the mid-eclipse times can also be determined. These timing measurements can be used to find more components and/or investigate a variety of different physical phenomena causing the orbital period changes of EBs \citep{2001icbs.book.....H,2001aocd.book.....K}.

EBs consisting of a pulsating member have attracted great scientific interest during the past two decades, as not only the stellar parameters of each star could be obtained in binaries, but also the internal structure of the pulsating stars could also be studied simultaneously with asteroseismology by exploring high-precision photometric observations. Among the pulsating EBs, about 208 systems have been classified as $\delta$ Sct type EB, which was also called as 'oscillating eclipsing Algol (oEA) stars' \citep{2004A&A...419.1015M,2017MNRAS.465.1181L,2018A&A...616A.130L}. $\delta$ Sct stars are short-period pulsators with period in the range of 0.02-0.25 day \citep{Breger2000}. They are located at the intersection of the main sequence and the classical Cepheid instability strip, and the spectral type is from A2 to F5 \citep{2001A&A...366..178R}. They usually pulsate in low-order radial/non-radial pressure (p) modes due to  $\kappa$ mechanism acting in the partial ionization of He II. These pulsations can be used to probe the envelope of a star with asteroseismology \citep{1999A&A...351..582H}. The $\delta$ Sct components in binaries share some similar features with single $\delta$ Sct stars, but the pulsations in oEA might be influenced by the companion, due to the mass transfer and gravitational forces from the companion \citep{2006MNRAS.366.1289S}. Recently, a threshold in the orbital period of about 13 days is found by \citet{2015ASPC..496..195L,2016arXiv160608638L}. Below this threshold, the pulsations would be influenced by the binarity. \citet{2013ApJ...777...77Z} investigated the relation between orbital and pulsation periods from theoretical perspective and found the pulsations in EBs depend on multiple factors, i.e. the system's orbital period, the mass ratio of the binary system, and filling factor of the primary pulsating member. Moreover, the tidal excited modes have also been found in eccentric-orbit binaries, and the resulting excited modes appear as the frequencies at multiples of the orbital frequency \citep{2011ApJS..197....4W,2013MNRAS.434..925H}.

Thanks to the ultra-high-precision photometric observations, $Kepler$ mission \citep{Borucki2010,Koch2010} has found more than 2878 eclipsing binary systems \citep{2016AJ....151...68K}, and at least 2000 $\delta$ Sct stars in the main field of view so far \citep{Balona2011,Balona2014,Bowman2016,Yang2018b,Yang2019,Yang2021}. Although many researchers have been hunting for and characterizing the pulsating EBs with $Kepler$ data, there are only about 30 new pulsating EBs with detailed investigation using $Kepler$ data \citep{2017MNRAS.470..915,2017MNRAS.465.1181L,2018A&A...616A.130L,2016MNRAS.460.4220L,2017ApJ...835..189L,2017ApJ...837..114G,2019AJ....157...17L}, and the interactions between the pulsations and orbital motion are still not fully understood. More pulsating EBs with detailed stellar parameters and investigations are needed.

KIC 8840638 ($\alpha_{2000}$=$19^{h}$$55^{m}$$35^{s}$.03, $\delta_{2000}$=+$45^{\circ}$$04^{'}$$45^{''}$.4, 2MASS: J19553503+4504454) was classified as a $\delta$ Scuti star by \citet{2014MNRAS.437..132R} according to its highest peak of the frequency spectra. Its pulsating period was about 49.6 min and this target was considered as a mid-late A type star in that survey \citep{2014MNRAS.437..132R}. Table \ref{tab:basic_parmeters} lists some basic parameters of KIC 8840638 collected from the survey and $Kepler$ Input Catalog (KIC; \citealt{Brown2011}). However, the light curve of KIC 8840638 shows obviously different features compared with that of a typical single $\delta$ Scuti star. In this paper, we investigate its light variation and astrophysical features using the high-precision $Kepler$ photometric time-series data. To avoid the Nyquist aliases detected in the frequency spectrum, only the SC data was used in this work.

\begin{table}
\begin{center}
\caption[]{Basic parameters of KIC 8840638.\label{tab:basic_parmeters}}
\begin{tabular}{lcc} \hline \hline
\noalign{\smallskip}
Parameters & KIC 8840638 &  References \\
\hline
\noalign{\smallskip}

 $K_P$          &  14.262   &   a  \\
 $P$            &  49.6 min &   b  \\
 $g_{SDSS}$     &  14.671   &   a  \\
 $i_{SDSS}$     &  14.086   &   a  \\
 $z_{SDSS}$     &  13.938   &   a  \\
 D51            &  14.489   &   a  \\
 $J_{2MASS}$    &  13.053   &   a  \\
 $H_{2MASS}$    &  12.708   &   a  \\
 $K_{2MASS}$    &  12.582   &   a  \\
 $g$            &  14.63    &   c  \\
 $U-g$          &   0.52    &   c  \\
 $g-r$          &   0.54    &   c  \\
 $T_{Gaia}$     &  5736 $\pm$ 250 K  &   d  \\
 $T_{KIC}$      &  6310 $\pm$ 250 K  &   a  \\
 $T_{GTC}$      &  7860 $\pm$ 120 K  &   b  \\
 log g          &  3.8  $\pm$ 0.25 dex & a  \\
\noalign{\smallskip}            
\hline   
\end{tabular}
\end{center}
\tablecomments{(a) KIC \citealt{Brown2011}; (b) \citealt{2014MNRAS.437..132R}; (c) \citealt{2012AJ....144...24G}; (d) https://gea.esac.esa.int/archive/.} 
\end{table}

\section{Observations and Data Reduction}

\begin{figure*}
  \centering
  \includegraphics[width=0.95\textwidth,trim=30 10 25 10,clip]{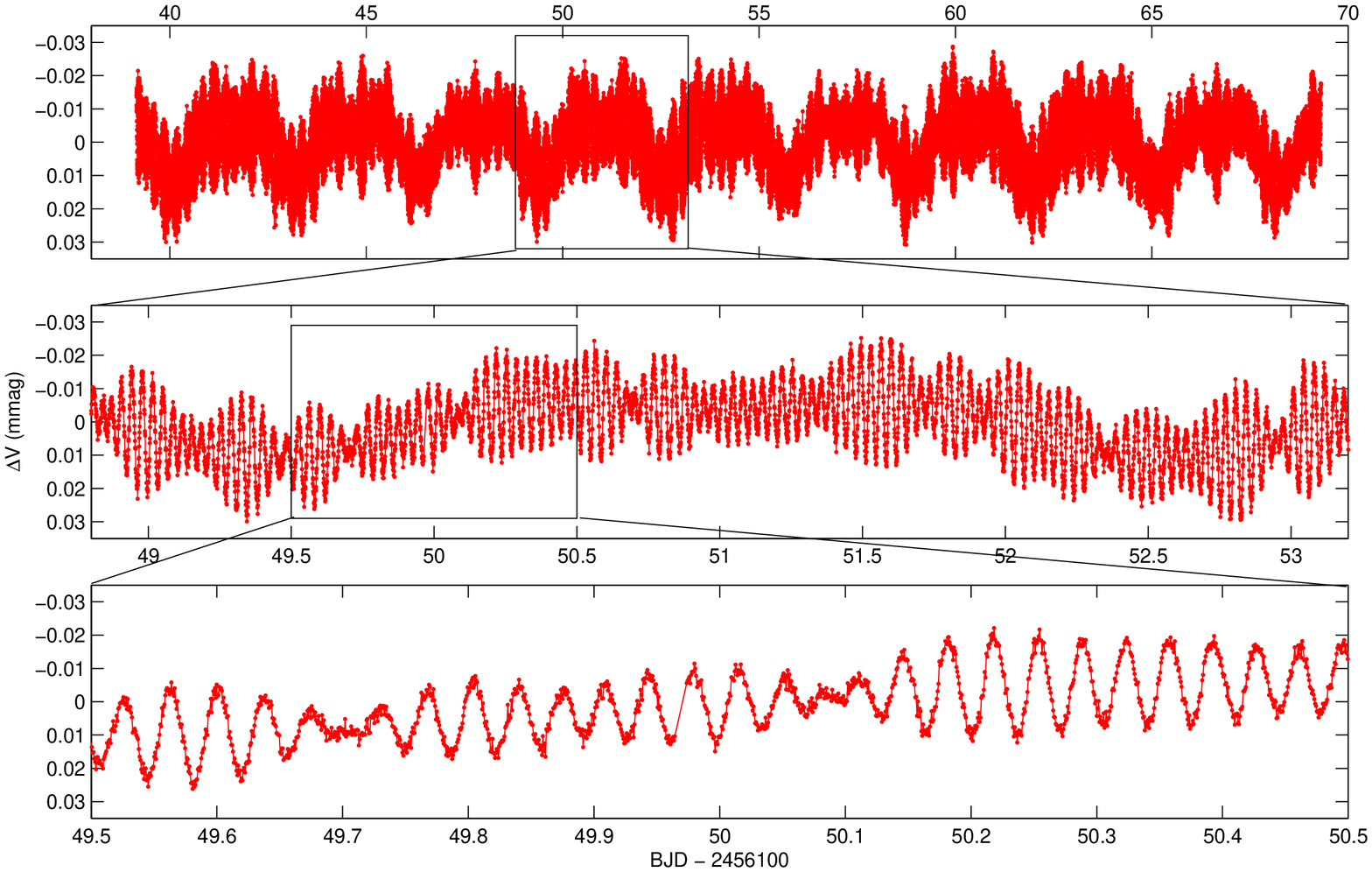}
\caption{The light curve of KIC 8840638 in short cadence. Top panel: the whole light curve of Q14.2. Middle panel: the zoom-in light curve of Q14.2 with 4.4 days. Bottom panel: the zoom-in light curve of Q14.2 with 1.0 day. }
\label{fig:SC_Q14_2_Three}
\end{figure*}

KIC 8840638 was observed by the $Kepler$ space telescope in short (integration time of 58.8 s) cadence from BJD 2456139.15 to 2456414.59, which spans 275.44 days. There are 4 quarters (Q14.2, Q15.2, Q16.2, and Q17.1) of the short cadence data, containing 165143 points in total. The short cadence photometric flux data of KIC 8840638 is available in Kepler Asteroseismic Science Operations Center (KASOC) data base\footnote{KASOC database: {http://kasoc.phys.au.dk}}\citep{Kjeldsen2010} with two types: the first is labeled as 'raw' data which was produced by the NASA Kepler Science pipeline, and the second is the flux data corrected by KASOC Working Group 4 (WG$\#$4: $\delta$ Scuti targets). We use the corrected data and perform corrections eliminating outliers, as well as the possible linear trends in some quarters. The flux data is converted to the magnitude scale, then the mean value of each quarter is subtracted, and the rectified time series is obtained. Figure \ref{fig:SC_Q14_2_Three} shows the rectified light curve of KIC 8840638 in SC data. It is clearly seen that the brightness of KIC 8840638 exhibits the feature of pulsation when one zooms in the light curve as shown in the middle and bottom panels of Figure \ref{fig:SC_Q14_2_Three}.

\section{Pulsational Characteristics}

To investigate the pulsating behavior, we used the software PERIOD04 \citep{Lenz2005} to perform Fourier analysis for the rectified data. To detect more significant frequencies in the SC data, we chose a frequency range of 0 $<$ $\nu$ $<$ 80 d$^{-1}$, a little wider than the typical pulsation frequency range of the $\delta$ Sct stars.

During the extraction of significant frequency, the highest peak in the frequency spectrum was considered as a significant frequency, then a multi-frequency least-squares fit using formula: $m = m_{0} + \Sigma\mathnormal{A}_{i}sin(2\pi(\mathnormal{f}_{i}\mathnormal{t} + \phi_{i}))$ ($m_{0}$ is the zero-point, $A_{i}$ is the amplitude, $f_{i}$ is the frequency, and $\phi_{i}$ is the corresponding phase) was conducted to the light curve with all the significant frequencies detected, resulting in the solutions of all the significant frequencies. A constructed light curve using the above solutions was subtracted from the data, and the residual was obtained to search for the next significant frequency. Then, the above steps were repeated until there was no significant peak in the frequency spectrum. The criterion (S/N $>$ 4.0) suggested by \citet{Breger1993} was adopted to judge the significant peaks. The uncertainties of frequencies were calculated following \citet{Kallinger2008}.

\begin{figure*}
\begin{center}
  \includegraphics[width=0.95\textwidth,trim=55 15 55 20,clip]{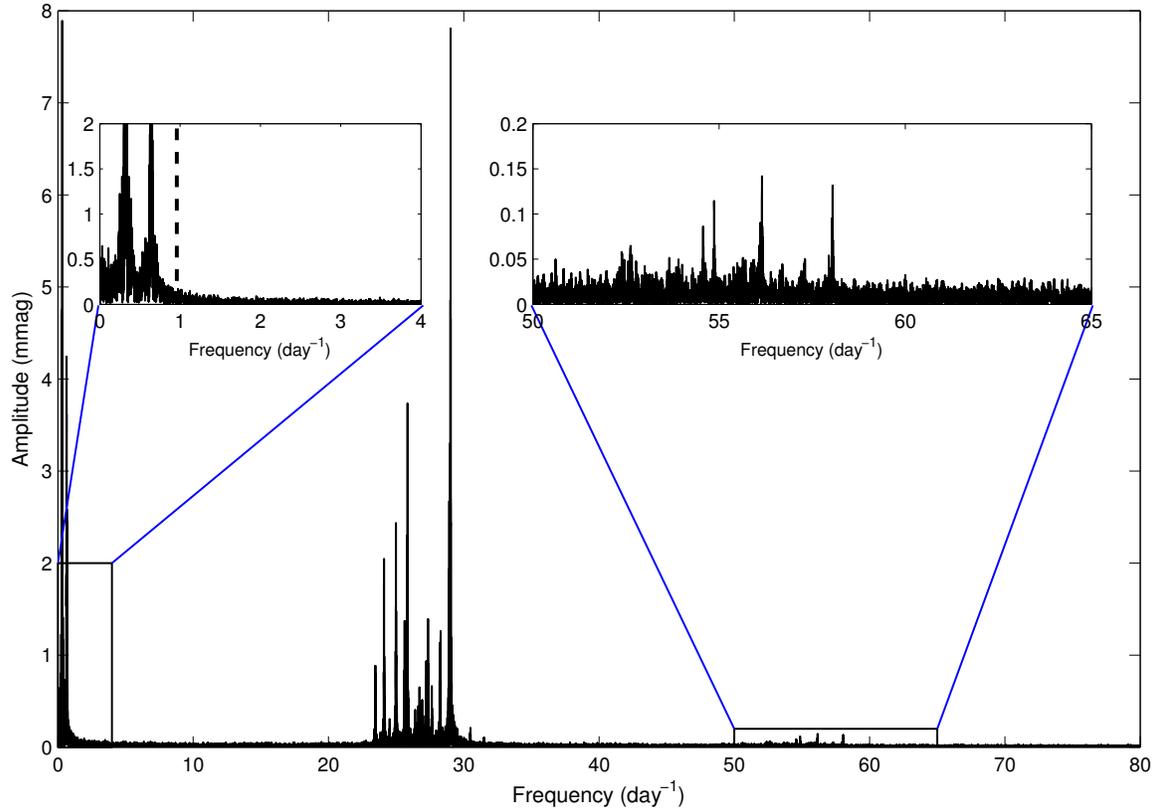}
  \caption{Amplitude spectra for the rectified SC data of KIC 8840638 up to 80 d$^{-1}$. The top-left insert figure: a zoom-in amplitude spectra in the low frequency region of 0 $<$ $\nu$ $<$ 4 d$^{-1}$. The top-right insert figure: a zoom-in amplitude spectra in the high frequency region of 50 $<$ $\nu$ $<$ 65 d$^{-1}$. }
    \label{fig:SC_spectra}
\end{center}
\end{figure*}

A total of 95 significant frequencies were extracted and they were listed in Table \ref{tab:Frequency-SC}. As shown in Figure \ref{fig:SC_spectra}, most of the detected frequencies lie in a frequency region between 23.0 and 32.0 d$^{-1}$. Among these frequencies, seven strong frequencies, i.e. $f_{2}$, $f_{4}$ to $f_{8}$, and  $f_{19}$ are considered to be independent frequencies, as they are neither any combinations nor harmonics of other frequencies, these seven frequencies are marked with 'independent' in the last column of Table \ref{tab:Frequency-SC}. Only some frequencies can be easily  identified as combinations with a simple form, and we denoted their identifications. For most of frequencies without identification, their forms of combination are very complicated, and hence may be just a coincidence.

In the low frequency region of 0 $<$ $\nu$ $<$ 1 d$^{-1}$, three frequencies, $f_{1}$(=0.320008 d$^{-1}$), $f_{3}$(=0.639945 d$^{-1}$) and $f_{62}$(=0.959744 d$^{-1}$) are interesting peaks. We examined these frequencies and found $f_{3}$ and $f_{62}$ are harmonics of $f_{1}$. The lowest frequency $f_{1}$ might be due to the orbital motion of binary. In the frequency region of 35 $<$ $\nu$ $<$ 80 d$^{-1}$, only several peaks were detected and they mainly concentrate in a narrow region of 52 $<$ $\nu$ $<$ 58 d$^{-1}$. These frequencies are combinations and harmonics of the independent terms and the orbital frequency.

\begin{deluxetable}{crcrcc}
\tabletypesize{\small}
\tablewidth{0pc}
\tablenum{2}
\tablecaption{All frequencies detected in SC data.\label{tab:Frequency-SC}}
\tablehead{
\colhead{$f_{i}$}   &
\colhead{Frequency (d$^{-1}$)}  &
\colhead{Amplitude (mmag)}      &
\colhead{S/N}            &
\colhead{Identification} &
}
\startdata
    1    &  0.320008 $\pm$ 0.000001 &   7.989   & 42.2 &  $f_{orb}$    &   \\
    2    & 29.024570 $\pm$ 0.000001 &   7.697   & 85.0 &  independent  &   \\
    3    &  0.639945 $\pm$ 0.000003 &   4.252   & 77.3 &  2$f_{orb}$   &   \\
    4    & 25.843653 $\pm$ 0.000003 &   3.780   & 36.3 &  independent  &   \\
    5    & 24.985826 $\pm$ 0.000005 &   2.441   & 30.3 &  independent  &   \\
    6    & 28.925876 $\pm$ 0.000005 &   2.205   & 35.0 &  independent  &   \\
    7    & 24.110686 $\pm$ 0.000006 &   2.032   & 45.6 &  independent  &   \\
    8    & 27.365182 $\pm$ 0.000008 &   1.408   & 13.5 &  independent  &   \\
    9    & 25.625751 $\pm$ 0.000008 &   1.401   & 39.0 &  f5+2$f_{orb}$  &   \\
    10   & 28.286089 $\pm$ 0.000009 &   1.218   & 18.8 &  f6-2$f_{orb}$  &  \\
    11   & 28.225151 $\pm$ 0.000011 &   1.074   & 17.8 &  f4+f8-f5  &  \\
    12   & 27.229833 $\pm$ 0.000012 &   0.944   & 12.7 &  2f6-f2-5$f_{orb}$  & \\
    13   & 23.470778 $\pm$ 0.000014 &   0.837   & 28.9 &  f7-2$f_{orb}$ & \\
    14   & 26.725303 $\pm$ 0.000018 &   0.652   &  9.5 &  f8-2$f_{orb}$  & \\
    15   & 27.636970 $\pm$ 0.000018 &   0.650   & 11.5 &   &  \\
    16   & 27.215143 $\pm$ 0.000018 &   0.651   & 12.8 &  f6+2f5-2f4  &  \\
    17   & 28.865007 $\pm$ 0.000021 &   0.541   & 16.2 &    &  \\
    18   & 26.934483 $\pm$ 0.000022 &   0.521   & 10.6 &    & \\
    19   & 24.099312 $\pm$ 0.000022 &   0.520   & 20.5 &  independent  &  \\
    20   & 27.254431 $\pm$ 0.000025 &   0.467   &  9.8 &    & \\
    21   & 26.637981 $\pm$ 0.000029 &   0.390   &  8.5 &    &  \\
    22   & 26.415518 $\pm$ 0.000031 &   0.369   &  9.0 &    &   \\
    23   & 26.997106 $\pm$ 0.000037 &   0.307   &  8.3 &    &   \\
    24   & 24.520008 $\pm$ 0.000040 &   0.290   & 14.7 &   &  \\
    25   & 23.459362 $\pm$ 0.000044 &   0.262   & 18.1 &  f4+f5-f8  &  \\
    26   & 27.205488 $\pm$ 0.000049 &   0.236   &  6.6 &    &   \\
    27   & 25.676200 $\pm$ 0.000055 &   0.211   &  8.0 &    &   \\
    28   & 30.494036 $\pm$ 0.000057 &   0.203   & 12.8 &    &   \\
    29   & 26.553998 $\pm$ 0.000058 &   0.198   &  6.0 &  f6+f7-f4-2$f_{orb}$ &  \\
    30   & 28.395747 $\pm$ 0.000064 &   0.181   &  6.1 &   & \\
    31   & 26.351489 $\pm$ 0.000068 &   0.170   &  5.7 &  f7+7$f_{orb}$  &  \\
    32   & 27.277842 $\pm$ 0.000069 &   0.167   &  5.0 &    &   \\
    33   & 23.845839 $\pm$ 0.000071 &   0.163   & 11.4 &   & \\
    34   & 25.890555 $\pm$ 0.000070 &   0.164   &  7.0 &    &   \\
    35   & 27.574478 $\pm$ 0.000072 &   0.160   &  5.0 &   & \\
    36   & 25.036422 $\pm$ 0.000077 &   0.150   &  8.5 &    &   \\
    37   & 27.055648 $\pm$ 0.000078 &   0.148   &  6.0 &    &   \\
    38   & 26.669568 $\pm$ 0.000078 &   0.147   &  5.5 &  f7+8$f_{orb}$     & \\
    39   & 27.955661 $\pm$ 0.000081 &   0.142   &  5.0 &    &   \\
    40   & 27.907678 $\pm$ 0.000077 &   0.149   &  5.7 &  f4+f8-f5-$f_{orb}$  &   \\
    41   & 27.208304 $\pm$ 0.000082 &   0.141   &  5.6 &   &   \\
    42   & 26.476246 $\pm$ 0.000082 &   0.140   &  6.3 &    &   \\
    43   & 58.049048 $\pm$ 0.000082 &   0.140   & 17.3 &  2f2  &   \\
    44   & 56.152527 $\pm$ 0.000088 &   0.130   & 10.5 &  f5+f6+7$f_{orb}$  &   \\
    45   & 28.715414 $\pm$ 0.000095 &   0.121   &  5.4 &    &   \\
    46   & 24.165708 $\pm$ 0.000094 &   0.122   &  8.9 &    &   \\
    47   & 27.316678 $\pm$ 0.000095 &   0.121   &  5.7 &    &   \\
    48   & 25.250567 $\pm$ 0.000096 &   0.120   &  7.5 &    &   \\
    49   & 54.868095 $\pm$ 0.000101 &   0.114   &  9.6 &  f2+f4  &   \\
    50   & 23.387329 $\pm$ 0.000101 &   0.114   & 11.1 &   & \\
    51   & 23.525758 $\pm$ 0.000106 &   0.108   & 11.3 &   &   \\
    52   & 27.309647 $\pm$ 0.000107 &   0.107   &  5.2 &  f7+10$f_{orb}$ &  \\
    53   & 27.550568 $\pm$ 0.000114 &   0.101   &  5.0 &  f5+8$f_{orb}$ &   \\
    54   & 28.670572 $\pm$ 0.000114 &   0.101   &  5.0 &   &   \\
    55   & 28.998015 $\pm$ 0.000117 &   0.098   &  5.9 &    &   \\
    56   & 27.109034 $\pm$ 0.000121 &   0.095   &  5.6 &    &   \\
    57   & 26.148543 $\pm$ 0.000125 &   0.092   &  5.2 &  f6+f5-f4-6$f_{orb}$  &   \\
    58   & 28.030988 $\pm$ 0.000128 &   0.090   &  4.4 &    & \\
    59   & 26.910598 $\pm$ 0.000129 &   0.089   &  5.3 &  f5+6$f_{orb}$  & \\
    60   & 26.089366 $\pm$ 0.000135 &   0.085   &  5.8 &  f8-4$f_{orb}$  & \\
    61   & 26.530479 $\pm$ 0.000135 &   0.085   &  5.2 &    & \\
    62   &  0.959744 $\pm$ 0.000140 &   0.082   &  4.1 &  3$f_{orb}$  & \\
    63   & 23.927733 $\pm$ 0.000140 &   0.082   &  7.7 &    & \\
    64   & 25.711425 $\pm$ 0.000140 &   0.082   &  5.7 &  f7+5$f_{orb}$  & \\
    65   & 28.293114 $\pm$ 0.000140 &   0.082   &  4.6 &  f6-2$f_{orb}$  & \\
    66   & 29.033747 $\pm$ 0.000146 &   0.079   &  4.2 &    & \\
    67   & 56.106080 $\pm$ 0.000149 &   0.077   &  6.6 &    & \\
    68   & 24.610641 $\pm$ 0.000158 &   0.073   &  5.9 &    & \\
    69   & 54.571444 $\pm$ 0.000164 &   0.070   &  6.2 &    & \\
    70   & 29.132862 $\pm$ 0.000164 &   0.070   &  4.2 &    & \\
    71   & 26.788611 $\pm$ 0.000169 &   0.068   &  4.4 &    & \\
    72   & 24.805965 $\pm$ 0.000169 &   0.068   &  5.5 &    & \\
    73   & 25.305529 $\pm$ 0.000177 &   0.065   &  5.5 &  f5+$f_{orb}$  & \\
    74   & 22.674102 $\pm$ 0.000198 &   0.058   &  7.8 &    & \\
    75   & 52.622924 $\pm$ 0.000202 &   0.057   &  5.4 &    & \\
    76   & 55.923463 $\pm$ 0.000205 &   0.056   &  4.9 &    & \\
    77   & 29.453077 $\pm$ 0.000205 &   0.056   &  4.1 &    & \\
    78   & 25.342924 $\pm$ 0.000209 &   0.055   &  4.8 &    & \\
    79   & 53.911772 $\pm$ 0.000209 &   0.055   &  4.9 &  f5+f6  & \\
    80   & 54.634252 $\pm$ 0.000221 &   0.052   &  4.7 &    & \\
    81   & 52.385486 $\pm$ 0.000225 &   0.051   &  5.0 &  f6+f4+f5-f8  & \\
    82   &  4.815395 $\pm$ 0.000225 &   0.051   &  6.7 &  f6-f7  & \\
    83   & 54.009898 $\pm$ 0.000230 &   0.050   &  4.7 &  f2+f5  & \\
    84   & 55.650909 $\pm$ 0.000235 &   0.049   &  4.5 &  f6+f8-2$f_{orb}$  & \\
    85   &  3.181020 $\pm$ 0.000235 &   0.049   &  5.0 &  f2-f4  & \\
    86   & 19.483597 $\pm$ 0.000235 &   0.049   &  6.9 &    & \\
    87   & 24.027238 $\pm$ 0.000235 &   0.049   &  5.4 &    & \\
    88   & 52.660710 $\pm$ 0.000250 &   0.046   &  4.3 &    & \\
    89   & 57.950430 $\pm$ 0.000256 &   0.045   &  5.0 &  f2+f6  & \\
    90   & 25.186838 $\pm$ 0.000256 &   0.045   &  4.2 &    & \\
    91   & 55.556078 $\pm$ 0.000267 &   0.043   &  4.1 &    & \\
    92   & 56.702608 $\pm$ 0.000274 &   0.042   &  4.3 &    & \\
    93   & 56.028777 $\pm$ 0.000274 &   0.042   &  4.0 &    & \\
    94   & 52.560506 $\pm$ 0.000274 &   0.042   &  4.4 &    & \\
    95   & 11.278539 $\pm$ 0.000280 &   0.041   &  5.6 &  3f12-3f13  & \\
    \enddata
\end{deluxetable}

\section{Binary Modeling}

\begin{figure} 
  \centering
  \includegraphics[width=1.05\columnwidth,angle=0]{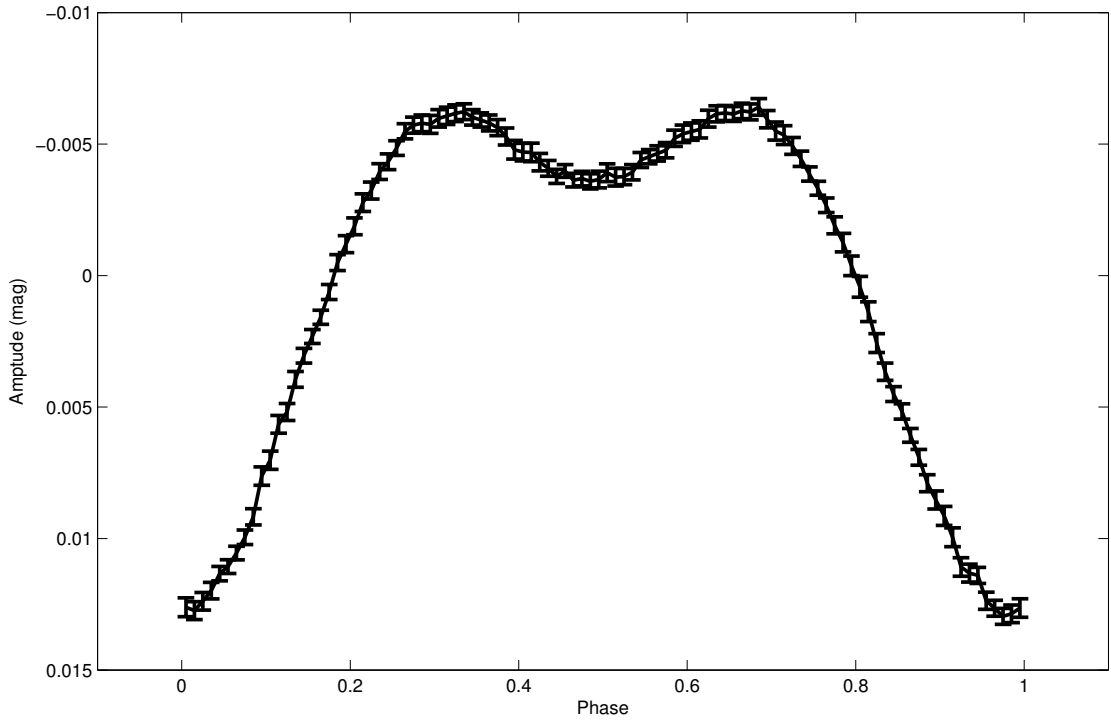}
  \caption{The folded orbital phase diagram of KIC 8840638 with the detected frequency $f_{1}$. For clearly, 100 time-bins are shown in an orbital phase.}
   \label{fig:SC_phase}
\end{figure}

As listed in Table \ref{tab:basic_parmeters}, it is clear that the temperature difference of KIC 8840638 collected from KIC and GTC is more than 1500 K. Such a large temperature difference in this star implies it may be a binary system. Moreover, from the top and middle panels of Figure \ref{fig:SC_Q14_2_Three}, the light variation of KIC 8840638 exhibits the typical shape of the light curve of eclipsing binaries, which suggests this target should belong to an eclipsing binary containing a pulsating star. To clearly show the light variation due to the orbital motion from binary, the light curve is folded with the detected low frequency of $f_{1}$. Figure \ref{fig:SC_phase} shows the phase diagram of the SC data, which includes 100 time-bins. From this figure, the orbital phase of KIC 8840638 is clearly shown, so the lowest frequency $f_{1}$ was marked as the orbital frequency (denoted as '$f_{orb}$') in Table \ref{tab:Frequency-SC}. It is obvious that the depth of the primary and secondary eclipses is different, which indicates a large temperature difference between the two components.

To obtain the physical and geometrical parameters of this binary system, we used the 2013 version of the Wilson-Devinney (W-D) code (\citealt{1971ApJ...166..605W};\citealt{1979ApJ...234.1054W}; \citealt{2012AJ....144...73W}) to analyze the rectified SC data. Considering the multiperiodic pulsations could affect the solution of the observed light curve, the removal of pulsating signals will be more conducive to the solution of binary parameters by W-D code. Thus, the pulsating signals in the light curve were smoothed out, only leaving the light variations due to the orbital motion from the binary.

The initial effective temperature of the hotter component is set to be 7860\,$\pm$\,120\,K from GTC while the other component is set to be 5736\,$\pm$\,250\,K which is released by Gaia2 \citep{2018A&A...616A...8A}. The reason for this setting is that the spectra of KIC 8840638 was assumed to contain the spectral components of the two stars. The temperatures derived from the spectrum are in fact the combination temperatures, which should vary with the rotation of the binary system. Different spectral temperatures can be set as the initial temperature of the two components in the light curve fitting process, e.g., Liu et al (2019) set the temperatures of different components based on different spectral temperatures at different phases. 

We defined parameters of the hotter component with subscript 1 and the cooler one with subscript 2. Based on the initial temperature of the two components of KIC 8840638, the gravity darkening coefficients and the bolometric albedos are fixed at standard values of $\textit{g}_{1}$ = 1.0, $\textit{g}_{2}$ = 0.32 \citep{1967ZA.....65...89L}, $\textit{A}_{1}$ = 1.0 and $\textit{A}_{2}$ = 0.5 \citep{1969AcA....19..245R}. The bolometric (X and Y) and the monochromatic (x and y) limbdarkening coefficients in logarithmic form are taken from \citet{1993AJ....106.2096V}.

\begin{figure*}
\begin{center}
  \includegraphics[width=0.46\textwidth,trim=0 40 5 15,clip]{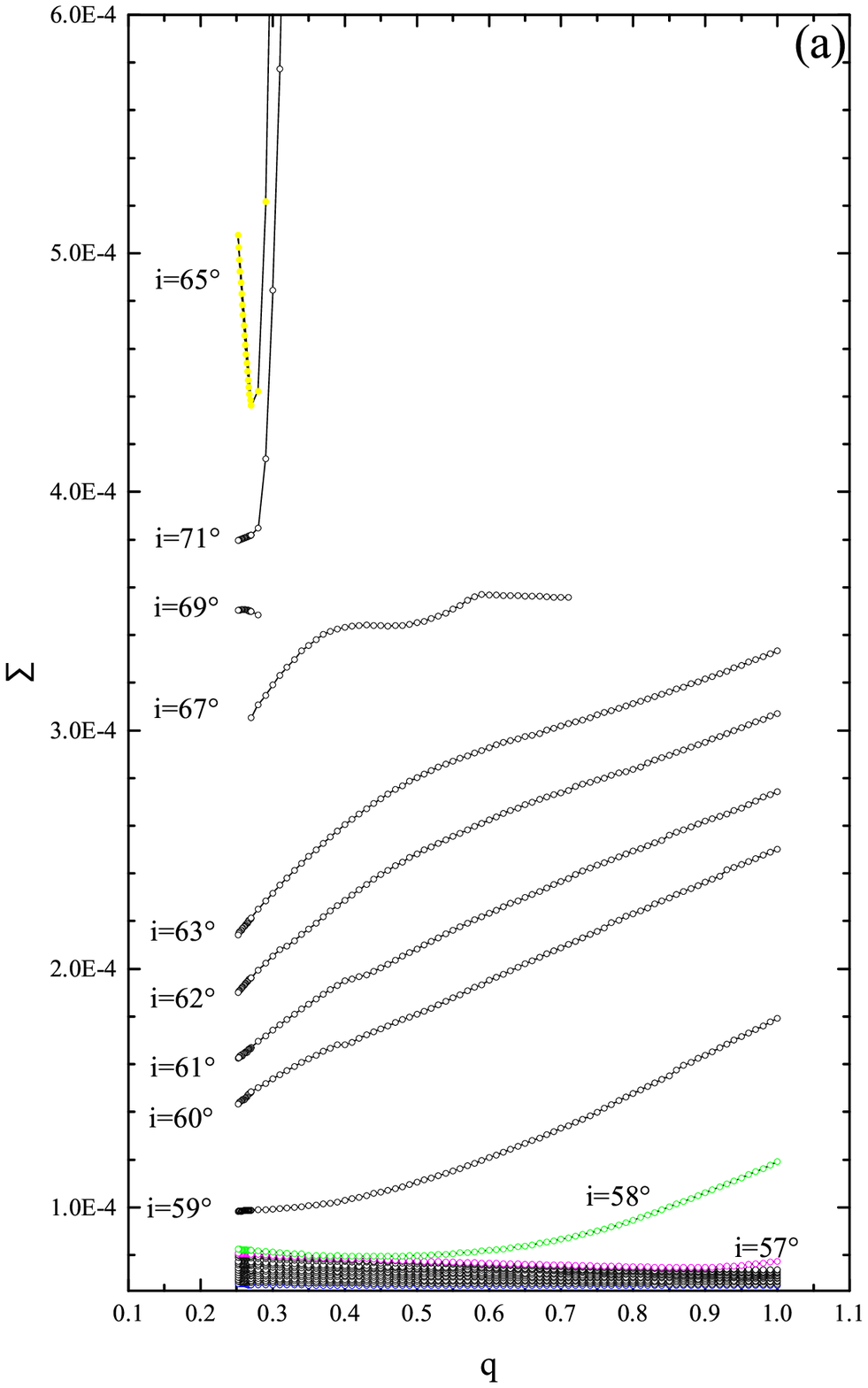}
  \includegraphics[width=0.46\textwidth,trim=0 40 5 15,clip]{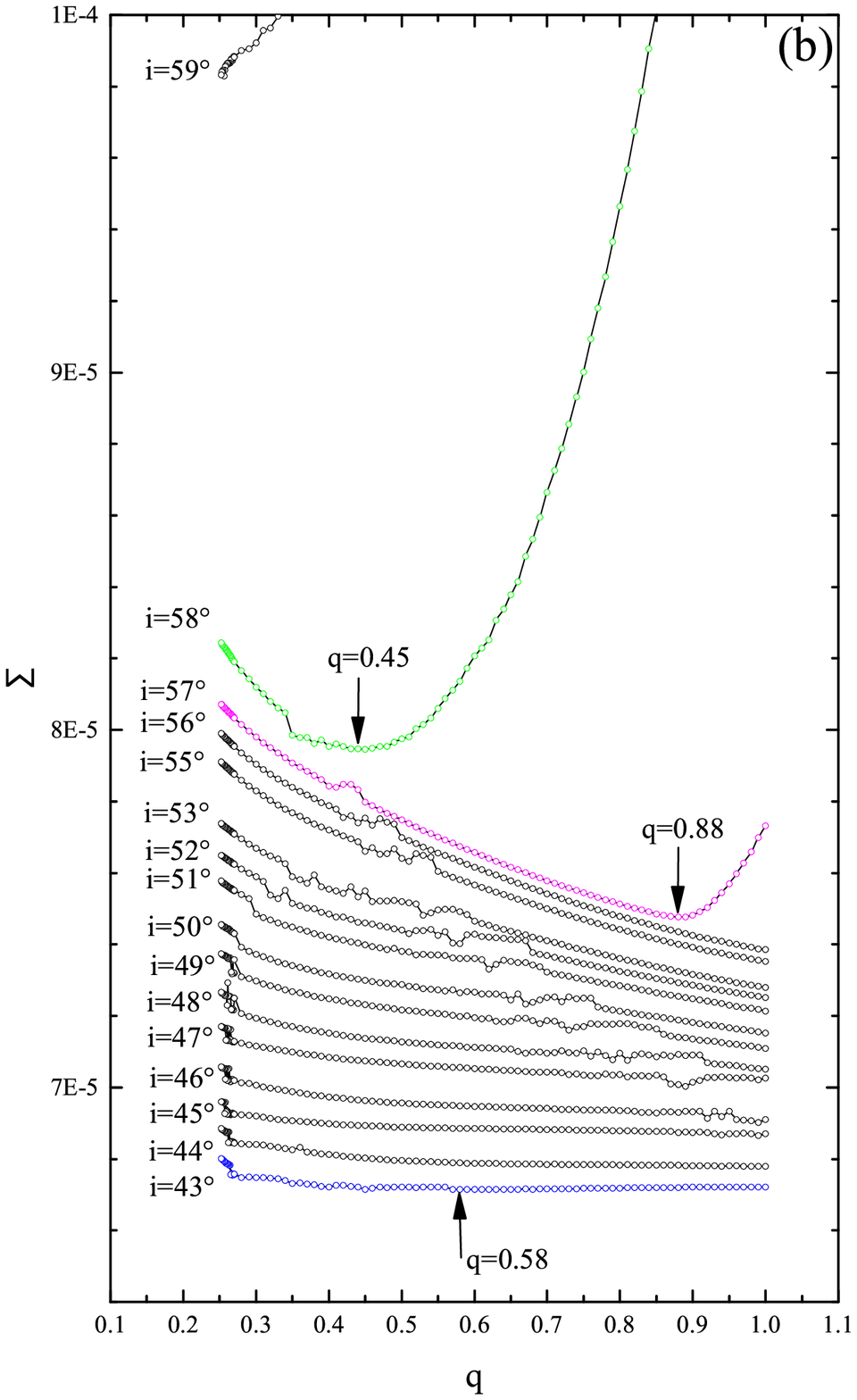}
  \caption{(a) Behavior of $\sum$ (the weighted sum of the residuals squared) of KIC 8840638 as a function of mass-ratio $q$ and inclination $i$. (b) Zoom in from $i$ = 43$^{\circ}$ to $i$ = 59$^{\circ}$ in the panel (a). For the curves of $i$ = 65$^{\circ}$ (yellow), 58$^{\circ}$ (green), 57$^{\circ}$ (magenta) and 43$^{\circ}$ (blue), they have minimal values at $q$ = 0.27, 0.45, 0.88 and 0.58, respectively. The open circles represent the q-search results at different inclinations for the semi-detached configurations.}
    \label{fig:q_search}
\end{center}
\end{figure*}

The mass ratio $q$ and orbital inclination $i$ are usually crucial parameters for the light curve analysis, however, these values are not available for KIC 8840638 at present, so we create a grid of $q$ and $i$ to search for the reliable values. During the search, the orbital inclination $i$ was set to vary from 1$^{\circ}$ to 90$^{\circ}$ with a step of 1$^{\circ}$. For each $i$, $q$ is fixed as a series of values (from 0.25 to 1.0 with a step of 0.01). The weighted sum of squared deviations $\sum(W(O-C)^{2})$ (hereafter $\sum$) of each convergent solution generated by W-D code was used to assess a potential reality of each combination of $q$ and $i$. The search process of $q$ and $i$ mentioned above was applied for various binary models in W-D code, however, the results show that only the Mode 5, which represents a semi-detached binary where the secondary component fills its limiting lobe, could derive an acceptable convergent solution. Figure \ref{fig:q_search} shows the behavior of $\sum$ of KIC 8840638 as a function of mass-ratio $q$ and inclination $i$. It should be noted that not all of the values in 1$^{\circ}$ $<$ $i$ $<$ 90$^{\circ}$ are present here, as other values lead to no convergent solution. Among all the values of $i$, only when $i$ is equal to 65$^{\circ}$ (abandoned due to the biggest $\sum$), 58$^{\circ}$, 57$^{\circ}$ and 43$^{\circ}$ with $q$ = 0.27, 0.45, 0.88 and 0.58 respectively, the $\sum$ curve possesses a minimal value. For other $i$ with value greater than 58$^{\circ}$, the value of $\sum$ shows a monotonically increasing trend; when it is less than 57$^{\circ}$, the value of $\sum$ shows a monotonically decreasing trend, which means the minimal value of $\sum$ is obtained in the interval where $q$ is greater than 1. From Figure \ref{fig:q_search}, we also note that the smallest minimal value of $\sum$ is achieved around $i$ = 43$^{\circ}$ and $q$ = 0.58, but in fact this is a locally minimal solution, so we do not use it as the initial value of light curve fitting. In addition, as shown in this figure, the value of $q$ is sensitive to different $i$, it is necessary to fix the value of $i$ in final light curve fitting. Hence, the value of $i$ was fixed separately as 58$^{\circ}$ and 57$^{\circ}$ with its corresponding $q$ as initial values.

\begin{figure*}
\begin{center}
  \includegraphics[width=1.0\columnwidth,trim=0 40 10 40,clip]{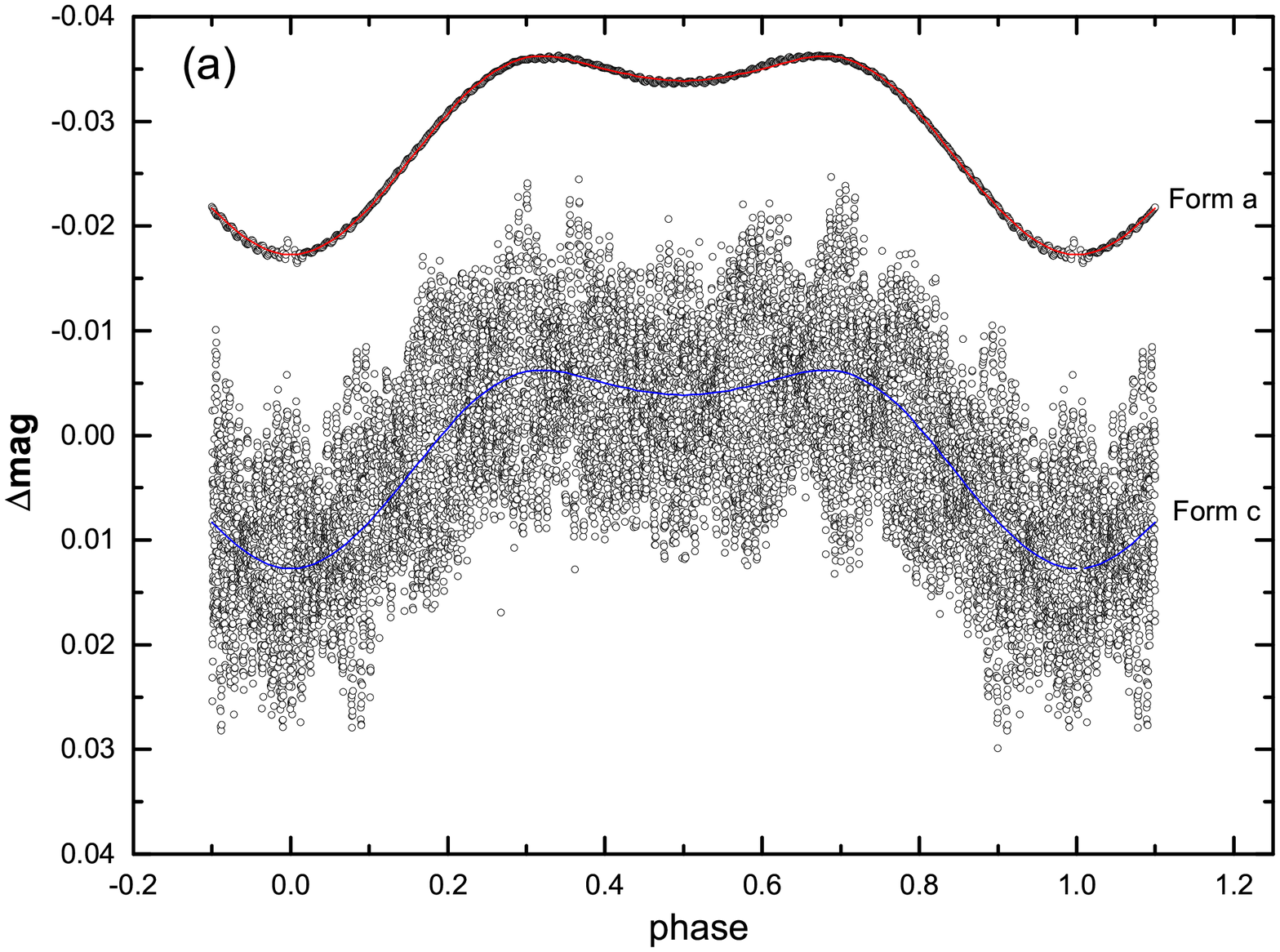}
  \includegraphics[width=1.0\columnwidth,trim=0 40 10 40,clip]{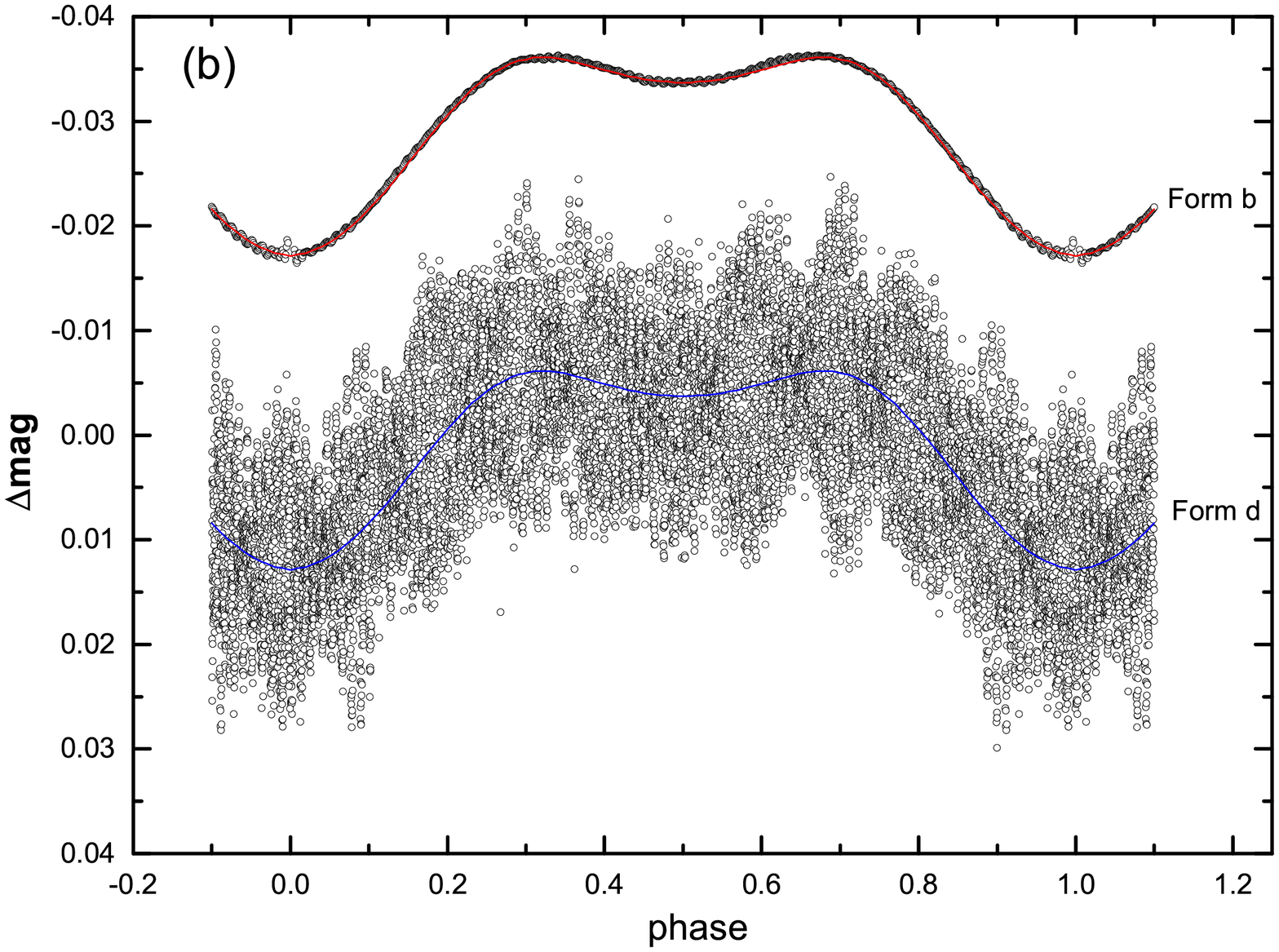}
  \caption{(a) Comparison of observed (open circles) and theoretical (solid lines) light curves at $i$ = 57$^{\circ}$. (b) Comparison of observed and theoretical light curves at $i$ = 58$^{\circ}$. In these two panels, the folded light curve with the pulsation removed is located on the upper part and the one without pulsation on the lower part. The theoretical light curves calculated from the parameter combinations of Form $a$, $b$, $c$ and $d$ in Table 1 is drawn with solid lines of the same name.}
    \label{fig:lightcurvefitting}
\end{center}
\end{figure*}

Then, the adjustable parameters are: the mean surface temperature of primary \textit{T}$_{1}$, the mean surface temperature of secondary \textit{T}$_{2}$, the mass ratio \textit{q}, the bandpass luminosity of primary \textit{L}$_{1}$, and the modified dimensionless surface potential, \textit{$\Omega$}$_{1}$ and \textit{$\Omega$}$_{2}$. The solutions derived from the phase-folded light curve, which excludes the pulsational variations, are listed as 'Form \textit{a}' ($i$ = 57$^{\circ}$) and 'Form \textit{b}' ($i$ = 58$^{\circ}$) in Table \ref{table:WD_solution}, and the synthetic light curves are shown in the upper parts of Figure \ref{fig:lightcurvefitting} (a) and (b).

To investigate the influence of pulsation on the orbital parameters of the binary system, the observed $Kepler$ light curve was also solved using W-D code. The solutions, given as 'Form \textit{c}'(\textit{i} = 57$^{\circ}$) and 'Form $d$' ($i$ = 58$^{\circ}$), are listed in the second and fourth columns of Table \ref{table:WD_solution}. From Table \ref{table:WD_solution}, we find the binary parameters from Form $a$ and $b$ are in good agreement with those from Form $c$ and $d$, respectively, which implies the photometric solutions of KIC 8840638 might not be influenced by the pulsation. Figure \ref{fig:lightcurvefitting} (a) and (b) show the synthetic light curves from these four forms respectively, they match well with the corresponding observed phase-folded light curves.

\begin{deluxetable}{lcccc}
\tabletypesize{\small}
\tablewidth{0pc}
\tablenum{5}
\tablecaption{Photometric solutions of KIC 8840638.\label{table:WD_solution}}
\tablehead{
Parameters &\underline{$~~$Form a$^{1}$}  &\underline{$~~$Form c$^{2}$} &\underline{$~~$Form b$^{1}$} & \underline{$~~$Form d$^{2}$} \\
 &Primary$~~$Secondary  &Primary$~~$Secondary  &Primary$~~$Secondary &Primary$~~~$Secondary \
}
\startdata
$g$ (deg)       & 1.00    $~$  0.50   & 1.00 $~$  0.50      & 1.00  $~$  0.50        & 1.00    $~$  0.50   \\
$A$ (deg)       & 0.50    $~$  0.32   & 0.50 $~$  0.32      & 0.50  $~$  0.32        & 0.50    $~$  0.32   \\
$X$             & 0.670   $~$  0.640  & 0.670 $~$ 0.640     & 0.670 $~$  0.640       & 0.670   $~$  0.640  \\
$Y$             & 0.200   $~$  0.170  & 0.200 $~$ 0.170     & 0.200 $~$  0.170       & 0.200   $~$  0.170 \\
$x$             & 0.585   $~$  0.715  & 0.585 $~$ 0.715     & 0.585 $~$  0.715       & 0.585   $~$  0.715  \\
$y$             & 0.253   $~$  0.212  & 0.253 $~$ 0.212     & 0.253 $~$  0.212       & 0.253   $~$  0.212 \\
$i$ (deg)       & 57.00               & 57.00               & 58.00                       & 58.00  \\
$q=M_{2}/M_{1}$ & 0.879(9)            & 0.870(14)           & 0.454(21)                   & 0.454(13)   \\
$T_{eff}$ (K)   & 7935(50) $~$ 2994(50)  & 7931(30) $~$ 2981(40)  & 7778(40) $~$ 3244(20)  & 7777(30)$~$  3244(30)\\
$\Omega_{in}$   & 3.55                   & 3.55                   & 2.78                  & 2.78    \\
$\Omega_{out}$  & 3.06                   & 3.06                   & 2.51                  & 2.51    \\
$\Omega$        & 6.20(17)$~$ 3.55       & 6.20(12)$~$  3.55      & 4.82(13)  $~$ 2.78    & 4.82(9) $~$ 2.78 \\
$L_{1}/(L_{1}+L_{2})$  &  0.986(6)       & 0.985(8)               & 0.987(9)              & 0.988(7) \\
$r$(pole)  & 0.1873(68)$~~$ 0.3454(34) & 0.1870(49)$~~$ 0.3445(52)  & 0.2282(46)$~~$ 0.2922(51) & 0.2282(52)$~~$ 0.2915(78) \\
$r$(side)  & 0.1885(33)$~~$ 0.3622(61) & 0.1882(51)$~~$ 0.3613(37)  & 0.2302(62)$~~$ 0.3049(39) & 0.2300(67)$~~$ 0.3041(81) \\
$r$(back)  & 0.1900(59)$~~$0.3936(92)  & 0.1897(84)$~~$ 0.3927(74)  & 0.2319(74)$~~$ 0.3374(82) & 0.2317(49)$~~$ 0.3367(63)  \\
$\sum W(O-C)^{2}$  & 0.000075      & 0.0028         & 0.000079         & 0.0027 \\

\enddata
\tablecomments{1: Result from the light curve removing the pulsations. 2: Result from the observed $Kepler$ light curve. }
\end{deluxetable}

Comparing the solutions of the light curve in Form $a$ and Form $b$, the stellar parameters in the former are slightly larger than that in the latter, but the values of $\sum$ are almost the same. In Form $a$, the larger mass ratio ($q$ = 0.88) means the physical parameters of the two stars (e.g. effective temperature $T_{eff}$) should be closer, however, the derived solution give a significant temperature difference. Among the solution in Form $b$, the temperatures of the two components are closer to the initial temperatures. Moreover, the a relatively small $q$ could be responsible for the temperature difference. Therefore, the solutions in Form $b$ seems to provide the more reasonable parameters for KIC 8840638. 

For the primary component of KIC 8840638, its effective temperature ($T$ = 7778 K) corresponds to a normal main-sequence star with a spectral type of about A7V according to the relationship by \citet{1988BAICz..39..329H}. Considering the uncertainties of the temperature given in Table \ref{table:WD_solution} are probably underestimated, we assumed the uncertainty of each component was 200 K. Using the empirical relation between spectral type and the stellar mass, the mass of the primary component was estimated to be $M_{1}$ = 1.75 $\pm$ 0.06 $M_{\odot}$, and the mass of the secondary component $M_{2}$ = 0.79 $\pm$ 0.03 $M_{\odot}$. Then the other absolute parameters of the primary and secondary component of KIC 8840638 can be derived from the photometric solutions and $M_{1}$ as follows: $R_{1}$ = 2.83 $\pm$ 0.12 $R_{\odot}$, log $g_{1}$ = 3.77 $\pm$ 0.06 (cgs), $L_{1}$ = 26 $\pm$ 5 $L_{\odot}$ and $R_{2}$ = 3.83 $\pm$ 0.13 $R_{\odot}$, log $g_{2}$ = 3.16 $\pm$ 0.07 (cgs), $L_{2}$ = 1.46 $\pm$ 0.52 $L_{\odot}$, respectively.

\section{DISCUSSION}

To investigate the pulsating behavior of KIC 8840638, we performed a multiple-frequency analysis using the high-precision  SC time-series data delivered from the $Kepler$ mission. Among the 95 detected significant frequencies, seven stronger terms are considered as the independent modes and they lie in the typical frequency range of $\delta$ Scuti stars. Three low frequencies ($f_{1}$=0.320008 d$^{-1}$, $f_{3}$=0.639945 d$^{-1}$ and $f_{62}$=0.959744 d$^{-1}$) are also detected directly in the frequency spectrum, and the lowest frequency $f_{1}$ was identified as the orbital frequency.

Based on 69 known eclipsing binaries containing $\delta$-Scuti-type components, \cite{2013ApJ...777...77Z} derived an upper limit of the $P_{pul} / P_{orb}$ ratio for the $\delta$-Scuti stars in eclipsing binaries with a value of 0.09, which could serve as a criterion to distinguish if a pulsating component in an eclipsing binary pulsates in the p-mode. For KIC 8840638, the period ratios of the above seven frequencies to orbital period were calculated to be $P_{pul} / P_{orb}$ = 0.011 $-$ 0.013, which were lower than the upper limit of 0.09 for $\delta$ Scuti stars in binaries. Hence, it seems that the seven independent frequencies belong to p modes of the $\delta$ Scuti stars.

To obtain the parameters of the binary system, we modelled the light curves using the W-D code. The light curves were satisfactorily modelled in two cases: including and removing the light variations due to the pulsations. The binary parameters from these two cases are consistent with each other, suggesting the binary parameters were not affected by the pulsations. From the light curve synthesis, this binary is a semi-detached EB system with $q$ = 0.454, an inclination angle of 58 degrees, and a temperature difference of larger than 4000 K between the components. In this semi-detached model, the primary component fills F1= 58 \% of their inner critical lobe and the secondary component fills full of their inner critical lobe. Here, the fill-out factor F1= $\Omega_{in}$ / $\Omega_{1}$, where $\Omega_{in}$ is the potential of the inner Roche surface. 

\begin{figure}
\begin{center}
  \includegraphics[width=1.0\columnwidth,trim=20 248 20 250,clip]{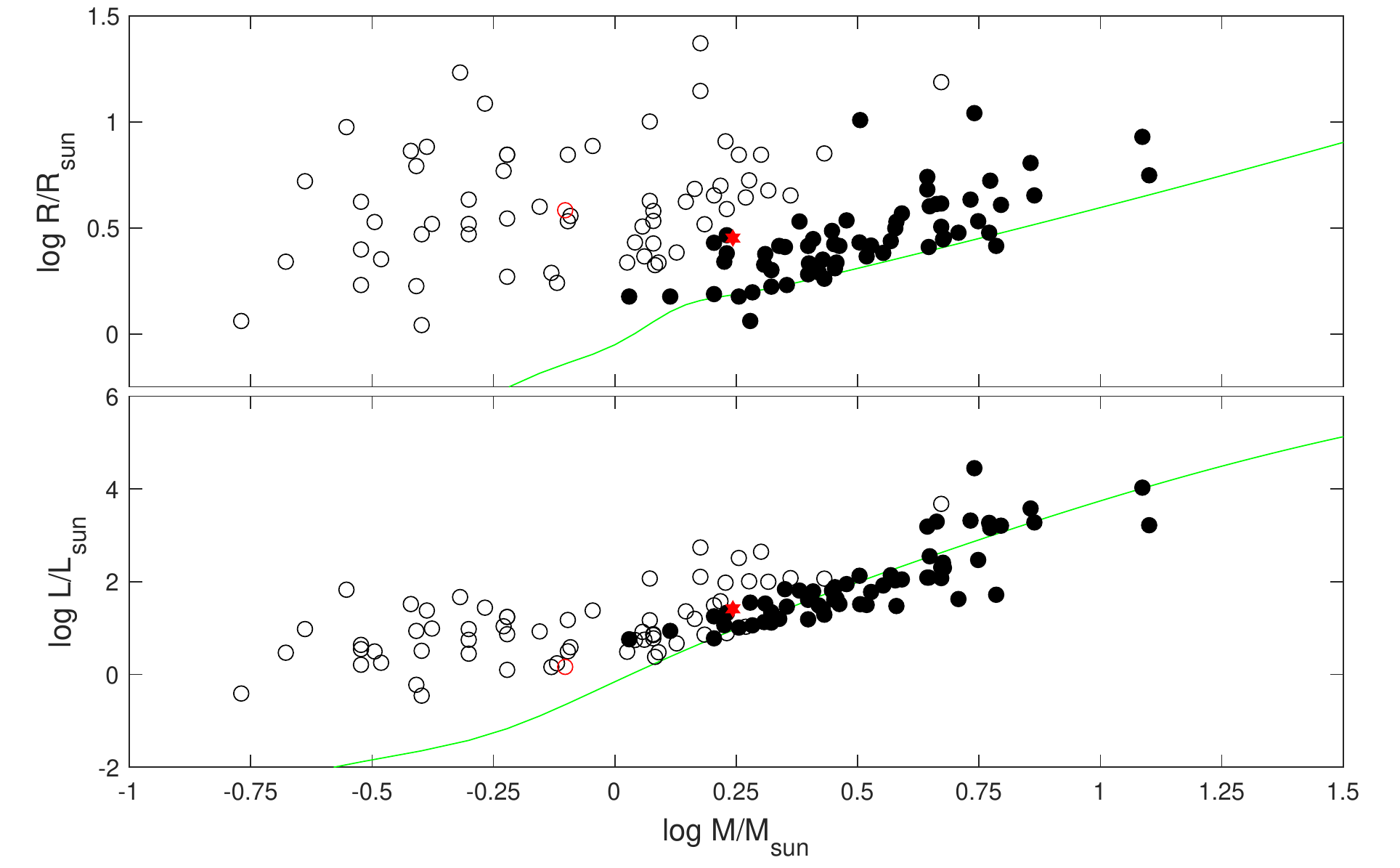}
  \caption{Locations of the primary ('red asterisk') and secondary component ('red circle') of KIC 8840638 in the mass-radius (top panel) and mass-luminosity (bottom panel) diagrams. The filled and open circles represent the primary and secondary components of the semi-detached Algol systems from \citet{2006MNRAS.373..435I} and \citet{2016MNRAS.460.4220L}. The green line represents the zero-age main-sequence (ZAMS) star from the model with a solar metallicity of Z = 0.02 provided by \citet{1996MNRAS.281..257T}.}
    \label{fig:Mass_R_L}
\end{center}
\end{figure}

\begin{figure}
\begin{center}
  \includegraphics[width=1.0\columnwidth,trim=43 240 48 265,clip]{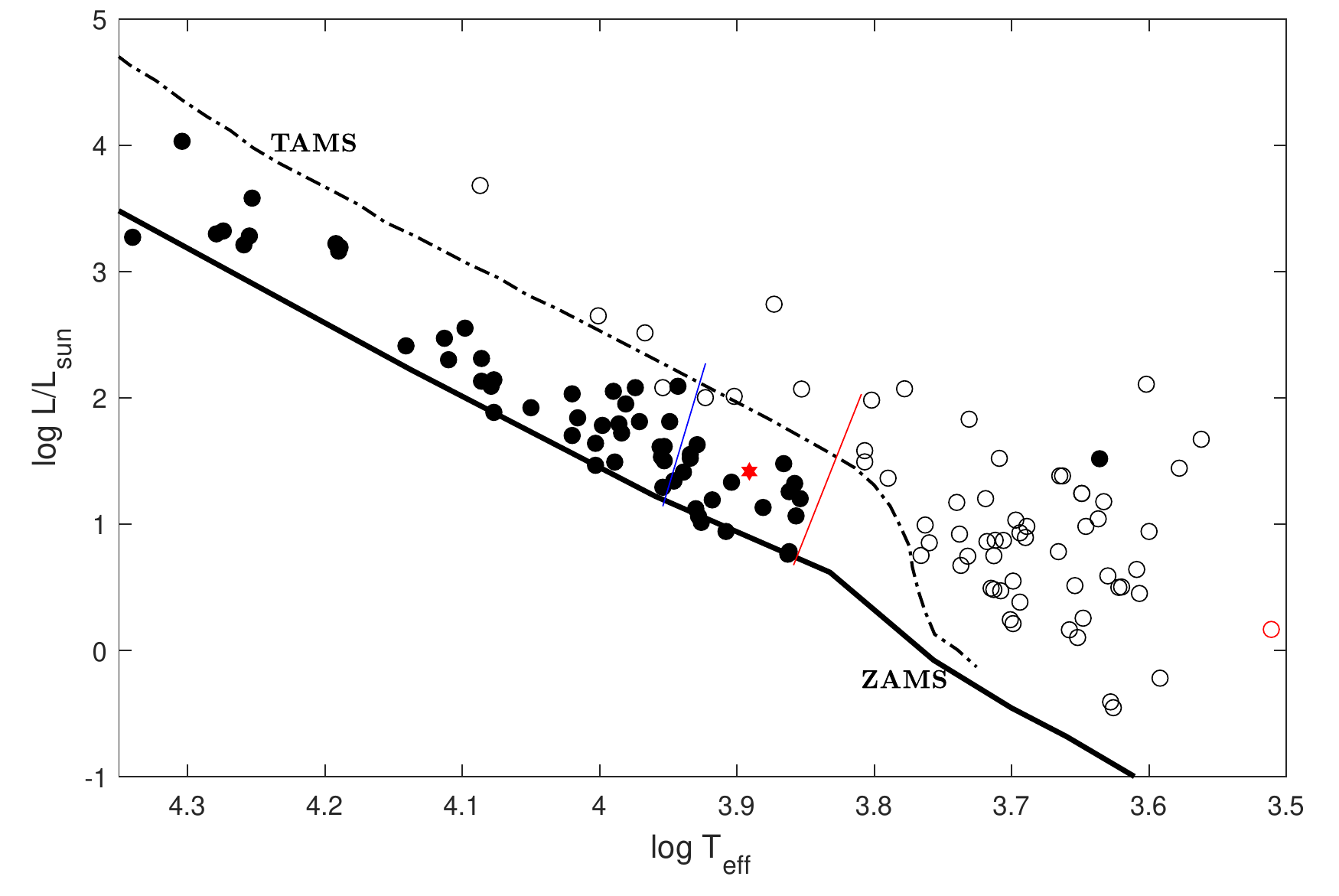}
  \caption{Locations of the primary ('red asterisk') and secondary component ('red circle') of KIC 8840638 in the H-R diagram. The filled and open circles are the same as in Figure \ref{fig:Mass_R_L}. The zero-age main-sequence (ZAMS) and terminal-age main-sequence (TAMS) are taken from the models in \citet{1998MNRAS.298..525P}. The blue and red solid lines represent the blue and red edge of $\delta$ Scuti instability strip \citep{2006MNRAS.366.1289S}.}
    \label{fig:HR_D}
\end{center}
\end{figure}

In Figure \ref{fig:Mass_R_L} and \ref{fig:HR_D}, we plot the locations of the components of KIC 8840638 in the mass-radius, mass-luminosity and H-R diagrams, together with those of other well-studied semi-detached Algol binary systems collected from the literature \citep{2006MNRAS.373..435I,2016MNRAS.460.4220L}. From Figure \ref{fig:Mass_R_L}, the radius and the luminosity of the primary star are close to the normal main sequence stars, which implies the primary component might be a main sequence star. The secondary star is also located in the general pattern of the semi-detached Algol systems, suggesting it evolves from the main sequence. Figure \ref{fig:HR_D} clearly shows the location of the primary components of KIC 8840638 lying in the $\delta$ Scuti instability region and below the terminal-age main-sequence (TAMS), suggesting it is likely a main sequence star. 

For the secondary component, although located in the general region of the secondary components of the semi-detached Algol systems, it is cooler than others and seems to be the coolest companion star among the semi-detached Algol systems. Moreover, compared with the normal dwarf stars of the same mass, the radius of the secondary star is about five times oversized and the luminosity is more than four times overluminous. These features suggest it evolves further than the others and seems to step into a highly evolved stage.  

\section{CONCLUSIONS}

In this paper, we investigated the light variation of KIC 8840638 using the Kepler short-cadence time-series data obtained from Quarter 14 to 17. The analysis of the light curve suggests this target is a new pulsating EB system, rather than a single pulsating star previously known. With the Fourier transformation of the light curve, we detected 95 significant frequencies, and seven of them are identified as independent p modes in the frequency range of 23.0 $-$ 32.0 d$^{-1}$. 

From the modellings of the $Kepler$ light curve, we found the pulsations were not influenced by the orbital motion. The synthetic light curves suggest this binary is a semi-detached EB system with $q$ = 0.454, an inclination angle of 58 degrees, and a temperature difference of larger than 4000 K between the components. With the derived effective temperatures and luminosities, the primary component of KIC 8840638 is considered as a main-sequence star with spectral type about A7V, while the secondary star is cooler than all the other companion stars of the semi-detached Algol systems, suggesting it steps into a highly evolved stage. 

As KIC 8840638 is a faint star, further radial velocity measurements from 8-10 meter class optical telescopes are needed to derive its masses more accurately, which may help us understand the physical properties and evolutionary stage of this system better.

\acknowledgements

This research is supported by the National Natural Science Foundation of China (grant Nos. 11573021, U1938104, 12003020) and the Fundamental Research Funds for the Central Universities. We would like to thank the $Kepler$ science team for providing such excellent data.

\end{document}